# Weak Energy: Form and Function

Allen D. Parks

Electromagnetic and Sensor Systems Department

Naval Surface Warfare Center

Dahlgren, VA 22448

allen.parks@navy.mil

**Abstract.** The equation of motion for a time-dependent weak value of a quantum mechanical observable contains a complex valued energy factor – the *weak energy of evolution*. This quantity is defined by the dynamics of the pre-selected and post-selected states which specify the observable's weak value. It is shown that this energy: (i) is manifested as dynamical and geometric phases that govern the evolution of the weak value during the measurement process; (ii) satisfies the Euler-Lagrange equations when expressed in terms of Pancharatnam (P) phase and Fubini-Study (FS) metric distance; (iii) provides for a PFS stationary action principle for quantum state evolution; (iv) time translates correlation amplitudes; (v) generalizes the temporal persistence of state normalization; and (vi) obeys a time-energy uncertainty relation. A similar complex valued quantity – the *pointed weak energy* of an evolving quantum state – is also defined and several of its properties in PFS coordinates are discussed. It is shown that the imaginary part of the pointed weak energy governs the state's survival probability and its real part is – to within a sign - the Mukunda-Simon geometric phase for arbitrary evolutions or the Aharonov-Anandan (AA) geometric phase for cyclic evolutions. Pointed weak energy gauge transformations and the PFS 1-form are defined and discussed and the relationship between the PFS 1-form and the AA connection 1-form is established.

## Preamble

It is an honor and a pleasure to speak at Yakir's 80[th] birthday conference. Although I first met Yakir about ten years ago, I have been a disciple of his work for many more years than that. I commemorate Yakir's birthday today by reviewing certain aspects of my research which have been inspired by Yakir's insights into foundational quantum physics via weak measurements and weak value theory. In particular, I will discuss the notion of weak energy – its relationships to both weak value measurements and quantum state evolution, as well as some of the formalism associated with it when it is expressed in terms of Pancharatnam phase and Fubini-Study metric distance.

## Weak Measurements and Weak Values

The *weak value $A_w$* of a quantum mechanical observable $A$ was introduced by Aharonov *et al* a quarter century ago [1-3]. It is the statistical result of a standard measurement procedure performed upon a pre-selected and post-selected (PPS) ensemble of quantum systems when the interaction between the measurement apparatus and each system is sufficiently weak, i.e. when it is a *weak measurement*. Unlike a strong measurement of $A$ which significantly disturbs the measured system (i.e., it "collapses" the wavefunction), a weak measurement of $A$ does not appreciably disturb the quantum system and yields $A_w$ as the observable's measured value. The peculiar nature of the virtually undisturbed quantum reality that exists between the boundaries defined by the PPS states is revealed by the eccentric characteristics of $A_w$, namely that $A_w$ is complex valued and that the real Re$A_w$ and imaginary Im$A_w$ parts of $A_w$ can be extremely large and lie far outside the eigenvalue spectral limits of $\hat{A}$. Although the interpretation of weak values remains somewhat controversial, experiments have verified several of the interesting unusual properties predicted by weak value theory [4 – 10].

Weak values arise in the von Neumann description of a quantum measurement at time $t_0$ of a time independent observable $A$ that describes a quantum system in an initial pre-selected state $|\psi_i\rangle$ at $t_0$ and a final post-selected state $|\psi_f\rangle$ at $t_0$. If the initial normalized measurement pointer state is $|\phi\rangle$, then the pointer state immediately after the measurement is

$$|\Psi\rangle = \langle\psi_f|e^{-\frac{i}{\hbar}\gamma\hat{A}\hat{p}}|\psi_i\rangle|\phi\rangle,$$

where $\gamma$ is the measurement interaction strength and $\hat{p}$ is the pointer momentum operator conjugate to the pointer position operator $\hat{q}$. When the measurement is weak, i.e., when $\gamma$ is sufficiently small and the pointer uncertainty $\Delta q$ is much larger than $\hat{A}$'s eigenvalue separation (qualifications required for weak measurements are discussed in [5, 11]), then the last equation becomes

$$|\Psi\rangle \approx \langle\psi_f|\psi_i\rangle\hat{S}(\gamma A_w)|\phi\rangle.$$

Here

$$A_w = \frac{\langle \psi_f | \hat{A} | \psi_i \rangle}{\langle \psi_f | \psi_i \rangle}, \quad \langle \psi_f | \psi_i \rangle \neq 0, \tag{1}$$

is the weak value of observable $A$ and $\hat{S}(\gamma A_w) = e^{-\frac{i}{\hbar}\gamma A_w \hat{p}}$ is the translation operator for $|\phi\rangle$ defined by the action $\langle q | \hat{S}(\gamma A_w) | \phi \rangle = \phi(q - \gamma Re A_w)$. Assuming that $\phi(q)$ is real valued, then after the measurement the new mean pointer position and momentum are [12]

$$\langle \Psi | \hat{q} | \Psi \rangle = \langle \phi | \hat{q} | \phi \rangle + \gamma Re A_w$$

and

$$\langle \Psi | \hat{p} | \Psi \rangle = \langle \phi | \hat{p} | \phi \rangle + 2\gamma v(p) Im A_w,$$

where $v(p)$ is the variance of the initial pointer momentum.

## Weak Energy of Evolution

A fundamental aspect of weak value theory is the tenet that although the measurement of $\hat{A}$ occurs at time $t_0$, the PPS states appearing in (1) are actually pre-selected and post-selected at times $t_i < t_0$ and $t_f > t_0$, respectively. PPS states selected at these times define past and future boundary conditions which influence $A_w$ at measurement time $t_0$ via their unitary evolutions forward in time from $t_i$ to $t_0$ and backward in time from $t_f$ to $t_0$.

Let $T = [t_1, t_2]$ be a fixed closed time interval, $\Delta t_i$ and $\Delta t_f$ be fixed time intervals, and $\mathcal{A} = \{A_w(t): t \in T\}$ be a time ordered set, where (i) $A_w(t)$ is the weak value of $\hat{A}$ at a measurement interaction time $t \in T$ defined by a state $|\psi_i(t_i)\rangle$ which has been pre-selected at time $t_i = t - \Delta t_i$ and by a state $|\psi_f(t_f)\rangle$ which has been post-selected at time $t_f = t + \Delta t_f$; and (ii) these PPS states continuously change from their initial states at $t_i$ and $t_f$ according to

$$\frac{d|\psi_i(t_i)\rangle}{dt_i} = -\frac{i}{\hbar}\hat{H}_i|\psi_i(t_i)\rangle, \quad t_i \in [t_1 - \Delta t_i, t_2 - \Delta t_i],$$

and

$$\frac{d|\psi_f(t_f)\rangle}{dt_f} = -\frac{i}{\hbar}\hat{H}_f|\psi_f(t_f)\rangle, \quad t_f \in [t_1 + \Delta t_f, t_2 + \Delta t_f].$$

When the Hamiltonians $\hat{H}_i$ and $\hat{H}_f$ are non-vanishing and explicitly time independent, then the evolutions of these PPS states to the time of measurement $t$ are

$$|\psi_i(t)\rangle = e^{-\frac{i}{\hbar}\hat{H}_i \Delta t_i}|\psi_i(t_i)\rangle \equiv \hat{U}|\psi_i(t_i)\rangle$$

and

$$|\psi_f(t)\rangle = e^{\frac{i}{\hbar}\hat{H}_f \Delta t_f}|\psi_f(t_f)\rangle \equiv \hat{V}|\psi_f(t_f)\rangle$$

so that

$$A_w(t) = \frac{\langle\psi_f(t_f)|\hat{V}^\dagger \hat{A}\hat{U}|\psi_i(t_i)\rangle}{\langle\psi_f(t_f)|\hat{V}^\dagger \hat{U}|\psi_i(t_i)\rangle} = \frac{\langle\psi_f(t)|\hat{A}|\psi_i(t)\rangle}{\langle\psi_f(t)|\psi_i(t)\rangle}, \quad \langle\psi_f(t)|\psi_i(t)\rangle \neq 0. \quad (2)$$

Observe that since $[\hat{U}, \hat{H}_i] = 0 = [\hat{V}, \hat{H}_f]$, then the actions of $\hat{H}_i$ and $\hat{H}_f$ upon the PPS states at $t_i$ and $t_f$ are transformed into their actions upon the evolved PPS states at the measurement time $t$.

If $\dot{A}_w$ exists at each $t \in T$, then $A_w$ is a continuous function over $T$ and the equation of motion for $A_w(t)$ is obtained by taking the time derivative of (2):

$$\dot{A}_w = \frac{i}{\hbar}\left\{(H_f A - A H_i)_w - A_w(H_f - H_i)_w\right\}. \quad (3)$$

Here it is assumed that $\hat{A}$ is explicitly time independent. The peculiar complex valued factor $(H_f - H_i)_w$ appearing in the second term of (3) is the *weak energy of evolution* for the PPS system. This quantity is an artifact of the dynamics of the PPS states and is contemporaneous with the measurement time $t$.

Although $(H_f - H_i)_w$ is not directly measured during the measurement of $A_w(t)$, it defines phases and phase factors which are crucial to determining $A_w(t)$ at the measurement time. To see this, consider the general solution to (3) given by

$$A_w = e^{-\frac{i}{\hbar}\int_{t_1}^{t}(H_f - H_i)_w dt'}\left\{A_w(t_1) + \frac{i}{\hbar}\int_{t_1}^{t} e^{\frac{i}{\hbar}\int_{t_1}^{t'}(H_f - H_i)_w dt''}(H_f A - A H_i)_w dt'\right\}.$$

This solution clearly shows that the time integrated weak energy of evolution is an intrinsic attribute of $A_w(t)$ and that it determines and influences $A_w(t)$ through phase factors which have been introduced by the integrating factor $e^{\frac{i}{\hbar}\int_{t_1}^{t}(H_f - H_i)_w dt'}$.

The argument of this integrating factor is the phase sum [13]

$$\frac{1}{\hbar}\int_{t_1}^{t}(H_f - H_i)_w dt' = \delta_f(t) - \delta_i(t) + \beta_f(t) - \beta_i(t),$$

where

$$\delta_x(t) \equiv \frac{1}{\hbar}\int_{t_1}^{t}\langle\psi_x(t')|\hat{H}_x|\psi_x(t')\rangle dt', \quad x \in \{i,f\},$$

are real valued dynamical phases and

$$\beta_f(t) \equiv \frac{1}{\hbar}\int_{t_1}^{t}\Delta H_f \frac{\langle\psi_f^\perp(t')|\psi_i(t')\rangle}{\langle\psi_f(t')|\psi_i(t')\rangle} dt'$$

and

$$\beta_i(t) \equiv \frac{1}{\hbar}\int_{t_1}^{t}\Delta H_i \frac{\langle\psi_f(t')|\psi_i^\perp(t')\rangle}{\langle\psi_f(t')|\psi_i(t')\rangle} dt'$$

are complex valued geometric phases. Here, $|\psi_x^\perp(t)\rangle$ belongs to the Hilbert subspace which is orthogonal to the subspace containing $|\psi_x(t)\rangle$, $x \in \{i,f\}$, and satisfies $\langle\psi_x^\perp(t)|\psi_x(t)\rangle = 0$, as well as the energy uncertainty condition $\Delta H_x = \langle\psi_x^\perp(t)|\hat{H}_x|\psi_x(t)\rangle$.

That $\beta_f$ is a geometric phase follows from the facts that $\beta_f$ is invariant under (i) local $U(1)$ gauge transformations $|\psi_x(t')\rangle \to e^{i\theta(t')}|\psi_x(t')\rangle, x \in \{i,f\}$, (which implies $|\psi_f^\perp(t')\rangle \to e^{i\theta(t')}|\psi_f^\perp(t')\rangle$) and (ii) the reparameterization $|\psi_f(t')\rangle = |\psi_f'(\tau(t'))\rangle \equiv |\psi_f'(\tau)\rangle$ (which implies $|\psi_f^\perp(t')\rangle = |\psi_f^{\perp\prime}(\tau)\rangle$). Here $\tau(t')$ is monotone increasing over the interval $[\tau(t_1), \tau(t)]$ with endpoints $|\psi_f'(\tau(t_1))\rangle = |\psi_f(t_1)\rangle$ and $|\psi_f'(\tau(t))\rangle = |\psi_f(t)\rangle$ [14]. Similar arguments hold for $\beta_i$.

Thus, it may be concluded that *the time accumulation of the forward and backward time evolutions to the measurement time t of the actions of $\hat{H}_i$ and $\hat{H}_f$ upon the associated PPS states is physically manifested at t as the sum $\delta_f(t) - \delta_i(t) + \beta_f(t) - \beta_i(t)$ of dynamical and geometric phases which determines and influences $A_w$ at t via the associated exponential phase factors.*

# The Weak Energy of Evolution Stationary Action Principle

Let $\mathcal{H}$ be $A_w(t)$'s Hilbert space and $\mathcal{P}$ be the associated projective space consisting of all the rays of $\mathcal{H}$. Recall that a ray is an equivalence class $[\psi]$ of states $|\psi\rangle$ in $\mathcal{H}$ which differ only in phase. If $\Pi: \mathcal{H} \to \mathcal{P}$ is the projection map $\Pi(|\psi\rangle) = [\psi]$, then the evolutions of the PPS states which define $A_w(t)$ at any time $t \in T$ are represented by the two curves $|\psi_i(t)\rangle$ and $|\psi_f(t)\rangle$ in $\mathcal{H}$ such that their projections in $\mathcal{P}$ at any $t \in T$ are separated by the Fubini-Study (FS) metric distance $s$ defined by [15,16]

$$s^2 \equiv s^2(t) = 4(1 - |\langle\psi_f(t)|\psi_i(t)\rangle|^2). \tag{4}$$

Also, if at any $t \in T$ the state $|\psi_f(t)\rangle$ is parallel transported along the unique path in $\mathcal{H}$ that is the pre-image under $\Pi$ of the shortest geodesic joining $[\psi_f(t)]$ and $[\psi_i(t)]$ in $\mathcal{P}$, then the associated Pancharatnam (P) phase $\chi$ is given by [17]

$$\chi \equiv \chi(t) = arg \frac{\langle\psi_f(t)|\psi_i(t)\rangle}{|\langle\psi_f(t)|\psi_i(t)\rangle|}. \tag{5}$$

Rearrangement of the time derivatives of (4) and (5) reveals that

$$Re(H_f - H_i)_w = \hbar\dot{\chi}$$

and

$$Im(H_f - H_i)_w = \hbar\left(\frac{s}{4-s^2}\right)\dot{s}.$$

This yields the following important identity which expresses the weak energy of evolution in terms of PFS coordinates [18]:

$$\mathcal{L}(s; \dot{s}, \dot{\chi}) \equiv (H_f - H_i)_w = \hbar\dot{\chi} + i\hbar\left(\frac{s}{4-s^2}\right)\dot{s}.$$

Since $\mathcal{L} \equiv \mathcal{L}(\dot{s}; \dot{s}, \dot{\chi})$ satisfies the Euler-Lagrange equations

$$\frac{d}{dt}\left(\frac{\partial\mathcal{L}}{\partial\dot{x}}\right) = \frac{\partial\mathcal{L}}{\partial x}, \quad x \in \{\chi, s\}, \tag{6}$$

it is called the *PFS Lagrangian* and the first variation of the *weak energy of evolution action* $\int_{t_1}^{t_2} \mathcal{L} dt$ vanishes, i.e.

$$\delta \int_{t_1}^{t_2} \mathcal{L} dt = 0.$$

Thus, the following *weak energy of evolution stationary action principle* may be stated [19]:

*The actual paths followed in $\mathcal{H}$ by $|\psi_f(t)\rangle$ and $|\psi_i(t)\rangle$ between their endpoint states at $t_1$ and $t_2$ during the closed time interval $[t_1, t_2]$ are such that the weak energy of evolution action is stationary for all variations of $\chi$, $s$, and time that vanish at the interval endpoints $t_1$ and $t_2$.*

## Some Additional Properties of $\mathcal{L}$

It is easily determined that the equation of motion for the PPS correlation amplitude

$$\varphi_t \equiv \langle \psi_f(t) | \psi_i(t) \rangle, t \in T = [t_1, t_2],$$

is given by

$$\dot{\varphi}_t = \frac{i}{\hbar} \mathcal{L} \varphi_t$$

and that it has as its solution

$$\varphi_t = e^{\frac{i}{\hbar} \int_{t_1}^{t} \mathcal{L} dt'} \varphi_{t_1}. \tag{7}$$

Thus, $\mathcal{L}$ defines a complex valued exponential multiplication factor which translates correlation amplitudes in time by capturing and transferring from one amplitude to another the essence of the state dynamics in $\mathcal{H}$ via the associated changes in $\mathcal{P}$ and the phase acquired from parallel transport in $\mathcal{H}$.

The transition probability at $t \in T$ associated with the PPS states is the square modulus of (7). This yields the identity

$$|\varphi_t|^2 = e^{-\frac{2}{\hbar}\int_{t_1}^{t} Im\mathcal{L}dt'}|\varphi_{t_1}|^2$$

which shows that $Im\mathcal{L}$ defines an exponential multiplication factor which time translates correlation probabilities and generalizes the temporal persistence of state normalization (because when $i = f$, then $s = 0$ and $|\varphi_t|^2 = |\varphi_{t_1}|^2$).

Consider the equation of motion for $\mathcal{L}$ by setting $\hat{A} = \hat{H}_f - \hat{H}_i$ in (3). Then

$$\dot{\mathcal{L}} = \frac{i}{\hbar}\left\{\left(H_f^2 - 2H_f H_i + H_i^2\right)_w - \mathcal{L}^2\right\}.$$

When this derivative vanishes, then $\mathcal{L}$ is a constant of the motion and it serves as a good weak quantum number for the associated PPS system. If $\tau$ is the characteristic time for $\mathcal{L}$ to be changed by an amount equal to the width of its statistical distribution $\Delta_w\mathcal{L} \equiv |\Delta_w^2\mathcal{L}|^{\frac{1}{2}}$, i.e. the weak energy uncertainty [3], then

$$\tau \approx \frac{\Delta_w\mathcal{L}}{|\dot{\mathcal{L}}|},$$

where $\Delta_w^2\mathcal{L} = \left(\left(\hat{H}_f - \hat{H}_i\right)^2\right)_w - \left(\left(H_f - H_i\right)_w\right)^2$ is the weak variance of $\mathcal{L}$. This indicates that appreciable differences between correlation amplitudes should be expected only for times with differences significantly greater than $\tau$ (note that $\tau \to \infty$ when $\mathcal{L}$ is a good weak quantum number).

For the special case that $\hat{H}_i$ and $\hat{H}_f$ are mutual constants of the motion, i.e. $[\hat{H}_i, \hat{H}_f] = 0$ and $\frac{d\hat{H}_x}{dt} = 0, x \in \{i, f\}$, then

$$\dot{\mathcal{L}} = \frac{i}{\hbar}\Delta_w^2\mathcal{L}, \tag{8}$$

and changes in $\mathcal{L}$ are precisely due to its weak variance (clearly, in this case $\mathcal{L}$ is a constant of the motion when $\Delta_w^2\mathcal{L} = 0$). Since

$$\Delta_w\mathcal{L} = \hbar^{\frac{1}{2}}|\dot{\mathcal{L}}|^{\frac{1}{2}}, \tag{9}$$

then the following associated time-energy uncertainty relation for $\mathcal{L}$ is readily obtained from the substitution of the square of (9) into the previous expression for $\tau$:

$$\tau\Delta_w\mathcal{L} \approx \hbar.$$

As a final point of interest, note from (6) that the generalized momentum $p_\chi$ that is conjugate to the coordinate $\chi$ is a constant of the motion for any PPS correlation amplitude. More specifically, $\dot{p}_\chi = 0$ since

$$p_\chi \equiv \frac{\partial \mathcal{L}}{\partial \dot{\chi}} = \hbar.$$

This suggests the following novel definition for Planck's constant:

$$h \equiv 2\pi \frac{\partial \mathcal{L}}{\partial \dot{\chi}}, \text{ where } \mathcal{L} \text{ is the PFS Lagrangian associated with any time dependent PPS correlation amplitude } \varphi_t, t \in T.$$

# Pointed Weak Energy, Pointed Probability Current, and Quantum Geometric Phase

Let $|\psi(t)\rangle$ be a state which is evolving in $\mathcal{H}$ under the action of the Hamiltonian $\hat{H}$. If $|\psi(0)\rangle$ and $|\psi(t)\rangle$ are used as the initial and fixed final states, respectively, and $\hat{H}_f - \hat{H}_i$ is replaced by $\hat{H}$, then the above development - with slight modifications - generally applies for this special case, yielding

$$\varphi_{t,0} \equiv \langle \psi(t)|\psi(0)\rangle = e^{\frac{i}{\hbar}\int_0^t \mathcal{L}_0 dt'} = \frac{1}{2}\sqrt{4 - s_0^2}\, e^{i\chi_0}, \quad \langle \psi(t)|\psi(0)\rangle \neq 0, \tag{10}$$

as the *pointed correlation amplitude* and

$$\mathcal{L}_0 \equiv \mathcal{L}_0(s_0; \dot{\chi}_0, \dot{s}_0) \equiv \frac{\langle \psi(t)|\hat{H}|\psi(0)\rangle}{\langle \psi(t)|\psi(0)\rangle} = \hbar\dot{\chi}_0 + i\hbar\left(\frac{s_0}{4 - s_0^2}\right)\dot{s}_0 = p_{\chi_0}\dot{\chi}_0 + p_{s_0}\dot{s}_0$$

as the *pointed weak energy* which is also a Lagrangian. Type equation here.Here use is made of the fact that $p_x = \frac{\partial \mathcal{L}_0}{\partial \dot{x}}, x \in \{\chi_0, s_0\}$.

From (10) it is seen that the *pointed correlation probability* $\Pr(s)$ is the survival probability for the initial state $|\psi(0)\rangle$ given by

$$Pr(s_0) \equiv \varphi_{t,0}^* \varphi_{t,0} = \frac{1}{4}(4 - s_0^2) \tag{11}$$

(so that $\varphi_{t,0} = \sqrt{\Pr(s_0)}\, e^{i\chi_0}$) and that it defines the associated *pointed probability current* $C_0$ as

$$C_0 \equiv \frac{d\Pr(s_0)}{dt} = -\frac{1}{2} s_0 \dot{s}_0 \,.$$

It is clear from this that since $C_0$ satisfies the Euler-Lagrange equation

$$\frac{d}{dt}\left(\frac{\partial C_0}{\partial \dot{s}_0}\right) = \frac{\partial C_0}{\partial s_0}$$

the first variation $\delta J_0$ of the *pointed probability current action* $J_0 \equiv \int_0^{t_2} C_0\, dt$ vanishes. This is formalized as *the pointed probability current stationary action principle* which states that [20]:

*The actual evolutionary path followed in $\mathcal{H}$ by the state $|\psi(t)\rangle$ between the end points $|\psi(0)\rangle$ and $|\psi(t_2)\rangle$ is such that $J_0$ is stationary for all variations in $s_0$ and time that vanish at the endpoints.*

The action of $\hat{H}$ upon $|\psi(t)\rangle$ can be uniquely written as [3]

$$\hat{H}|\psi(t)\rangle = \langle H\rangle |\psi(t)\rangle + \Delta H\, |\psi^\perp(t)\rangle, \tag{12}$$

where $\langle H\rangle = \langle\psi(t)|\hat{H}|\psi(t)\rangle$, $\Delta H = \sqrt{\langle H^2\rangle - \langle H\rangle^2}$, and $|\psi^\perp(t)\rangle$ satisfies the conditions $\langle\psi^\perp(t)|\psi(t)\rangle = 0$ and $\Delta H = \langle\psi^\perp(t)|\hat{H}|\psi(t)\rangle$. The following equivalent definition for the pointed weak energy is obtained when the dual form of (12) is first used to form the scalar product with the state $|\psi(0)\rangle$ and then this product is divided by $\langle\psi(t)|\psi(0)\rangle \neq 0$:

$$\mathcal{L}_0 \equiv \langle H\rangle + \Delta H\, \frac{\langle\psi^\perp(t)|\psi(0)\rangle}{\langle\psi(t)|\psi(0)\rangle}\,.$$

Then

$$\varphi_{t,0} = e^{\frac{i}{\hbar}\int_0^t \mathcal{L}_0\, dt'} = e^{i(\delta_0 + \beta_0)},$$

where

$$\delta_0 \equiv \frac{1}{\hbar}\int_0^t \langle H\rangle\, dt'$$

is a real valued dynamical phase and

$$\beta_0 \equiv \frac{1}{\hbar}\int_0^t \Delta H\, \frac{\langle\psi^\perp(t')|\psi(0)\rangle}{\langle\psi(t')|\psi(0)\rangle}\, dt'$$

defines the complex valued *pointed phase* acquired by the system as a result of the evolution of $|\psi(t')\rangle$ over the interval $[0, t]$.

The pointed phase is invariant under the local $U(1)$ gauge transformation $|\psi(t)\rangle \to e^{i\theta(t)}|\psi(t)\rangle$ (which implies that $|\psi^\perp(t)\rangle \to e^{i\theta(t)}|\psi^\perp(t)\rangle$), as well as under the reparameterization $|\psi(t)\rangle = |\psi'(t'(t))\rangle$ (which implies that $|\psi^\perp(t)\rangle = |\psi^{\perp\prime}(t'(t))\rangle$) over the interval $[t'(0), t'(t)]$. Here $t'(t)$ is monotone increasing with state endpoints $|\psi'(t'(0))\rangle = |\psi(0)\rangle$ and $|\psi'(t'(t))\rangle = |\psi(t)\rangle$ [14]. These two invariance properties imply that $\beta_0$ is a geometric phase [21] – the *pointed geometric phase*.

Since $\delta_0$ is real valued, $\mathcal{L}_0$ and $\beta_0$ are complex valued, and

$$\frac{1}{\hbar}\int_0^t \mathcal{L}_0 dt' = \delta_0 + \beta_0,$$

then

$$Re\beta_0 = \frac{1}{\hbar}\int_0^t Re\mathcal{L}_0\, dt' - \delta_0 = \chi_0(t) - \delta_0$$

and

$$Im\beta_0 = \frac{1}{\hbar}\int_0^t Im\mathcal{L}_0\, dt' = \ln\frac{2}{\sqrt{4 - s_0^2(t)}}, \tag{13}$$

where use has been made of the fact that $\chi_0(0) = s_0(0) = 0$.

It can be inferred from (11) and (13) that since

$$e^{-Im\beta_0} = \sqrt{\Pr(s_0(t))},$$

then $Im\beta_0$ governs the survival probability for $|\psi(0)\rangle$ and comparison of the expression for $Re\beta_0$ with (4) in [22] identifies $-Re\beta_0$ as the Mukunda-Simon phase. For the special case that the evolution of $|\psi(t)\rangle$ is cyclic, comparison of the expression for $Re\beta_0$ with (3) in [23] reveals that $-Re\beta_0$ is Aharonov-Anandan phase.

## Some Additional Properties of $\mathcal{L}_0$

Consider the transformation

$$\mathcal{L}_0^\theta = \mathcal{L}_0 - \hbar\dot{\theta}(\chi_0, s_0), \tag{14}$$

where $\theta \equiv \theta(\chi_0, s_0)$ is a dimensionless function with continuous derivatives. Since transformations of this form applied to Lagrangians are referred to as gauge transformations in the classical mechanics literature (e.g. [24]), then (14) is called a *pointed weak energy gauge transformation*. From (10) it is easy to see that (14) corresponds to the local $U(1)$ gauge transformation $|\psi(t)\rangle \to e^{i\theta_t}|\psi(t)\rangle$, i.e.

$$\varphi_{t,0}^\theta = e^{\frac{i}{\hbar}\int_0^t \mathcal{L}_0^\theta dt'} = e^{\frac{i}{\hbar}\int_0^t \mathcal{L}_0 dt'} e^{-i[\theta]_0^t} = \varphi_{t,0} e^{-i(\theta_t - \theta_0)} = \langle\psi(t)|e^{-i\theta_t}e^{i\theta_0}|\psi(0)\rangle,$$

where $\theta_t \equiv \theta(\chi_0(t), s_0(t)) = \theta$. It is interesting to note from (14) that the *PFS 1-form* $\omega_0 \equiv \frac{1}{\hbar}\mathcal{L}_0 dt$ transforms as a $U(1)$ gauge potential, i.e.

$$\omega_0^\theta = \omega_0 - d\theta = \omega_0 + d(i\, lng),$$

where $g = e^{i\theta} \in U(1)$.

Let $|\psi_0\rangle$ be a distinguished state in $\mathcal{H}$, $\mathcal{H}_{\sim\perp} \equiv \{|\psi\rangle \in \mathcal{H} : \langle\psi|\psi_0\rangle \neq 0\}$, $\mathcal{R}$ be the set of real numbers, and define the map $\Psi_0: \mathcal{H}_{\sim\perp} \to \mathcal{R}\times\mathcal{R}$ by

$$\Psi_0(|\psi\rangle) \equiv \left(arg\frac{\langle\psi|\psi_0\rangle}{|\langle\psi|\psi_0\rangle|}, 2\sqrt{1-|\langle\psi|\psi_0\rangle|^2}\right) = (\chi_0, s_0).$$

Note that $\Psi_0$ provides an equivalence classification of the states in $\mathcal{H}_{\sim\perp}$. More specifically, $|\psi\rangle$ and $|\psi'\rangle$ are equivalent under $\Psi_0$, i.e. $|\psi\rangle \sim |\psi'\rangle$, when $\Psi_0(|\psi\rangle) = \Psi_0(|\psi'\rangle)$. *PFS configuration space* is the image set $\mathcal{B} \equiv im\Psi_0 = [0,2\pi) \times [0,2) \subset \mathcal{R}\times\mathcal{R}$.

If the map $\alpha: [0,\tau] \to \mathcal{H}_{\sim\perp}$ defines an evolutionary path with $|\psi(t)\rangle = \alpha(t)$, then the composition of maps $\rho \equiv \Psi_0 \circ \alpha$ is the curve in $\mathcal{B}$ which describes this evolution and has $\rho(0) = (\chi_0(0), s_0(0)) = (0,0)$ as its *first point* and $\rho(\tau) = (\chi_0(\tau), s_0(\tau))$ as its *last point*. The curve $\rho$ is *simple* if $\rho$ is injective and is *closed* if $\rho(0) = \rho(\tau)$. The closed curve $\rho$ is simple if the restriction of $\rho$ to the domain $(0,\tau)$ is injective. An evolutionary path in $\mathcal{H}$ over $[0,\tau]$ for which $|\psi(t)\rangle \in \mathcal{H}_{\sim\perp}$, $t \in [0,\tau]$, and for which $\rho$ is a smooth curve is said to be *proper* and $\rho$ is the *proper evolution* in $\mathcal{B}$. Observe that $\omega_0$ is exact since $\omega_0 = df$, $f = \chi_0 - \frac{i}{2}\ln(\frac{4-s_0^2}{4})$, so that for any proper evolution $\rho$ in $\mathcal{B}$ which connects $(0,0)$ to any other point $(\chi_0', s_0')$,

$$\int_\rho \omega_0 = f(\chi_0', s_0'). \tag{15}$$

Thus, *the value of $\int_\rho \omega_0$ is independent of the path taken between $(0,0)$ and $(\chi_0', s_0')$ in $\mathcal{B}$.*

Now suppose that $|\psi(t)\rangle$ and $|\psi'(t)\rangle$ are proper evolutionary paths over $[0, \tau]$ in $\mathcal{H}$ such that $|\psi(0)\rangle = |\psi'(0)\rangle$, $|\psi(\tau)\rangle \sim |\psi'(\tau)\rangle$, i.e. they are *last point equivalent paths*, and $\rho$ and $\rho'$ are their respective paths in $\mathcal{B}$. If the same transformation $e^{i\theta}$ is applied to both of these states, then

$$\int_\rho \omega_0^\theta - \int_{\rho'} \omega_0^\theta = \int_\rho \omega_0 - \int_{\rho'} \omega_0$$

or

$$\int_C \omega_0^\theta = \int_C \omega_0.$$

Here use has been made of the facts that negative signs preceding line integrals over curves reverse the orientation of the curves and that $C$ is the simple smooth curve that is the union of $\rho$ and $-\rho'$. It can therefore be concluded that *for any last point equivalent paths in $\mathcal{H}$ the value of $\omega_0$ on $C$ (i.e. $\int_C \omega_0$) is invariant under pointed weak energy gauge transformations*. Furthermore, because of path independence and since $\rho$ and $\rho'$ have the same first and last points, then for any two last point equivalent paths in $\mathcal{H}$

$$\int_\rho \omega_0 = \int_{\rho'} \omega_0$$

so that

$$\int_C \omega_0 = 0.$$

Let $\Omega$ be a simple closed curve in the projective space $\mathcal{P}$ of $\mathcal{H}$ over the time interval $[0, \tau]$ with $[\psi(0)] = [\psi(\tau)]$ and such that its image in $\mathcal{B}$ is smooth and simple. When the lift of $\Omega$ is closed, then $|\psi(0)\rangle = |\psi(\tau)\rangle$ and the image in $\mathcal{B}$ is a loop with $(0,0)$ as its first and last point and which intersects the $\chi_0$ and $s_0$ axes only at the origin. However, if the lift of $\Omega$ is not closed, then $|\psi(t)\rangle = e^{i\lambda}|\psi(0)\rangle$ and the image of the lift in $\mathcal{B}$ is a path $\sigma$ which has $(\chi_0(\tau), 0) = (\lambda, 0)$ as its last point. It the follows from (15) that

$$\int_\sigma \omega_0 = f(\lambda, 0) = \lambda.$$

For the special case that the lift is a horizontal lift, then $\lambda$ is the geometric phase and

$$\int_\sigma \omega_0 = \int_\Omega \mathcal{A},$$

where $\mathcal{A}$ is the associated Aharonov-Anandan connection 1-form [25]. It can therefore be concluded that *for each such horizontal lift of an Ω in 𝓟 there exists a path σ in 𝓑 for which the last equation is satisfied*.

## Acknowledgment

The preparation of this contribution to the Festschrift commemorating Yakir Aharonov's eightieth birthday was supported by a grant from the Naval Innovation in Science and Engineering program sponsored by the Naval Surface Warfare Center Dahlgren Division.